\RequirePackage{fix-cm}
\documentclass[smallextended]{svjour3}       
\smartqed  
\usepackage{amssymb}
\usepackage{amsmath}
\usepackage{color}
\usepackage{graphicx}
\usepackage[justification=centering]{caption}
\usepackage{multirow}
\usepackage{extarrows}
\begin{document}

\title{Multiparty Quantum Sealed-Bid Auction Using Single Photons as Message Carrier
}


\author{Wen-Jie Liu     \and
        Hai-Bin Wang    \and
        Gong-Lin Yuan   \and
        Yong Xu         \and
        Zhen-Yu Chen    \and
        Xing-Xing An  \and
        Fu-Gao Ji     \and
        Gnim Tchalim Gnitou      
}


\institute{
           W.-J. Liu   \and
           H.-B. Wang
           \at Jiangsu Engineering Center of Network Monitoring, Nanjing University of Information Science \& Technology, Nanjing 210044, P.R.China
           \\
           \email{wenjiel@163.com}
           \and
           W.-J. Liu   \and
           H.-B Wang   \and
           Y. Xu      \and
           Z.-Y. Chen     \and
           G. T. Gnitou   \and
           \at School of Computer and Software, Nanjing University of Information Science \& Technology, Nanjing 210044, P.R.China
            \and
           G.-L. Yuan
           \at College of Mathematics and Information Science, Guangxi University, Nanning, Guangxi, P.R. China
            \and
           X.-X. An  \and
           F.-G. Ji
           \at School of Electronics \& Information Engineering, Nanjing University of Information Science \& Technology, Nanjing 210044, P.R.China
 }

\date{Received: December 31, 2014 / Accepted: date}

\maketitle

\begin{abstract}
In this study, a novel multiparty quantum sealed-bid auction protocol using the single photons as the message carrier of bids is proposed, followed by an example of three-party auction. Compared with those protocols based on the entangled states (GHZ state, EPR pairs, etc.), the present protocol is more economic and feasible within present technology. In order to guarantee the security and the fairness of the auction, the decoy photon checking technique and an improved post-confirmation mechanism with EPR pairs are introduced respectively.

\end{abstract}

\section{Introduction}
\label{intro1}
Due to the principles of quantum mechanics, such as no-cloning theorem, uncertainty principle, and entanglement characteristics, quantum system shows a lot of advantages than classic ones. In the past decades, quantum mechanics has been applied to various fields, including quantum key distribution (QKD)[1,2], quantum secret sharing [3,4], quantum secure direct communication [5-7], quantum private comparison [8-12], quantum voting [13, 14] and quantum sealed-bid auction (QSA).

The auction is one of the basic businesses in e-commerce, and it has wide potential application in some complex distributed systems, such as social network[15], cloud computing[16,17], and wireless sensor networks[18,19]. Usually, the auction consists of one auctioneer and several bidders, all bidders submit their own bids to the auctioneer, and then the auctioneer will determine who is the winner. A secure auction must guarantee the privacy of every bid's message and the fairness among the bidders. To be specific, (1) any bid cannot be known by others before the auctioneer publishes it; (2) the auctioneer cannot help any bidder to win the auction through changing his/her bid.

In 2009, Naseri [20] proposed the first quantum sealed-bid auction protocol based on GHZ states, in which the GHZ states were also utilized to check the security of the quantum channel. However, Qin et al. [21] and Yang et al. [22] respectively pointed out that the malicious bidder can launch the double controlled-NOT attack or use fake entangled particles to cheat the other's secret bids, and then they fixed the problem by adding decoy photons besides GHZ states for checking eavesdropping. At the same time, other remarks and improvements on Naseri's protocol have also been reported [23,24]. In the above QSA protocols [20-23], the auctioneer is considered to be honest. But, in practice a malicious bidder may collude with a dishonest auctioneer to win the auction, which results in damaging the fairness of the auction. In order to solve this problem, in 2010, Zhao et al. [25] improved Naseri et al.'s protocol [20] by introducing a post-confirmation mechanism using the single photons. Since then, other researchers [26-28] further improved the post-confirmation mechanism to resist some special collusion attacks among the bidders or between the bidders and the auctioneer. On the other hand, in 2010, Wang [29] proposed a new QSA protocol based on a set of order EPR pairs, in which, the two legitimate bidders can encrypt their respective secret message through performing two unitary operations, and deduce the opposite party's bid after the auctioneer received all the particles and finished the Bell measurement. In 2014, Liu et al. [30] found that in Wang's protocol a malicious bidder can launch the Twice-CNOT attack to obtain the other's bid, or the dishonest auctioneer may collude with one bidder and help him/her win the auction by changing his/her bid. To resist these two attacks, they applied a hash function method into a post-confirmation mechanism and adopted a QKD key to encrypt the bids.

All above QSA protocols [20-30] are based on the entangled states (GHZ state or EPR pairs). To the best of our knowledge, no such QSA protocol utilizing the single photons has been reported. As we all know, the GHZ state or EPR pairs is more difficult to be prepared in experiment than the single photon. In this study, we take the single photons as quantum resource, and propose a novel economic and feasible multiparty quantum sealed-bid auction (MQSA) protocol among $N$ parties. In the present protocol, the decoy photon checking technique is adopted to guarantee the privacy of every bid's message, i.e., a malicious bidder or a group of malicious bidders cannot get other's bid. In addition, we improved the post-confirmation mechanism based on EPR pairs by adding a permutation operator before the EPR sequence is transmitted, which is used to ensure the fairness of the auction.In comparison with the previous protocols [20-30], this protocol has its own other advantages except that single photon is easier to be prepared than GHZ state or EPR pairs. The auctioneer is only required to carry out the simpler single-particle measurement instead of the Bell-basis or GHZ-basis measurement. The bids can be directly encrypted into single-particles without being encoded by QKD in advance. The security checking of the quantum channel only uses the decoy photon checking technique without consuming any GHZ states, thus saves the consumption of quantum resources.

The rest of this paper is organized as follows: the whole process of the present MQSA protocol among $N$ parties is presented in Section 2. For convenience, an example of the three-party auction is given in Section 3. Finally, the security and efficiency are analyzed respectively in Section 4, following the discussion and inclusion in the last section.

\section{Multiparty quantum sealed-bid action protocol using single photons as message carrier}
\label{sec:2}
Suppose the auction consists of one auctioneer (Alice), and $N$-1 bidders (Bob, Charlie, ..., and Zach). Without loss of generality, suppose the bidders' binary messages have the same length, and the form of messages is $l^{m}_j=\{l^{1}_j,l^{2}_j,l^{3}_j,...,l^{m}_j\}$, where $l^{m}_j$ denotes the $m$-th binary bit of the message of the $j$-th bidder, $m$ presents the length of the binary messages, and $l^{m}_j\in\{0,1\}, j=1,2,...,N-1$. The whole procedures of our MQSA protocol are depicted as below,

\textbf{Step 1}. All parties agree that bidders can perform the two unitary operations $I$, $i\sigma_y$ to encode one bit classic information[31], where:
\begin{equation}\label{1}
\begin{split}
  I=&|0\rangle\langle0|+|1\rangle\langle1|,\\
  i\sigma_y=&|0\rangle\langle1|-|1\rangle\langle0|.
\end{split}
\end{equation}

\textbf{Step 2}. The auctioneer Alice generates $N-1$ groups $m$-length qubits randomly in one of $\{|0\rangle,|1\rangle,|+\rangle,|-\rangle\}$, and noted them as $P_j (j=1,2,...,N-1)$. At the same time, Alice selects $K\times(N-2)$ ($K$ is the detection rate) decoy photons from $\{|+\rangle,|-\rangle,|+y\rangle,|-y\rangle\}$ and inserts randomly $k$ decoy photons into $P_j$, respectively,
\begin{equation}\label{2}
\begin{split}
|+\rangle=&\frac{1}{\sqrt{2}}(|0\rangle+|1\rangle),|-\rangle=\frac{1}{\sqrt{2}}(|0\rangle-|1\rangle),\\
|+y\rangle=&\frac{1}{\sqrt{2}}(|0\rangle+i|1\rangle),|-y\rangle=\frac{1}{\sqrt{2}}(|0\rangle-i|1\rangle).
\end{split}
\end{equation}
Now, $P_j$ is transformed into a new sequence $P^{'}_j$. Alice sends $P^{'}_j$ to every bidder $j$.

\textbf{Step 3}. All bidders confirm that they have received sequence $P^{'}_j$, then, Alice and bidder $j$ perform the eavesdropper checking of quantum channel as follows. (i) Alice announces the positions and the corresponding MB (in the two sets of measuring basis (MB), say $\{|+\rangle,|-\rangle\}$ and $\{|+y\rangle,|-y\rangle\}$) of the decoy photons in $P^{'}_j$. (ii) Bidder $j$ measures decoy photons in sequence $P^{'}_j$ according to the information published by Alice. (iii) Bidder $j$ announces the measurement results. (vi) Through comparing the measurement results in public, they complete the error rate analysis. If the error rate is zero, they continue to next step, otherwise they abandon their transmission.

\textbf{Step 4}. After checking the security of quantum channel, the leftover qubits are the sequence $P_j$. Each bidder encrypts his bid $l_j$ on $P_j$ according to this rule: if $l^{m}_j$ is 0, the bidder performs $I$ operation on $P^{m}_j$, otherwise, the bidder performs $i\sigma_y$ operation on $P^{m}_j$. The sequence $P_j$ is transformed into sequence $P^{''}_j$.

\textbf{Step 5}. For $1\leq j\leq N-1$, bidder $j$ prepares ordered sequences of EPR pairs $R^{'}_{ji}=\{r^{1}_{ji},r^{2}_{ji},...,r^{m/2}_{ji}\}(i=1,2,...,N-1,j\neq i)$ according to his (her) secret bid, where $r^{g}_{ji}\in\{|\psi^{+}\rangle,|\psi^{-}\rangle,|\varphi^{+}\rangle,|\varphi^{-}\rangle\}(g=1,2,...,m/2)$ and that classical bits ``00'', ``01'', ``10'', ``11'' are encoded to states $|\psi^{+}\rangle,|\psi^{-}\rangle,|\varphi^{+}\rangle$, and $|\varphi^{-}\rangle$, respectively. Then, bidder $j$ applies permutation operators $\Pi^{j}_{(m)}$ on $R^{'}_{ji}$ to create a new sequence $\Pi^{j}_{(m)}R^{'}_{ji}=R^{''}_{ji}$ and transforms the sequence $R^{''}_{ji}$ into new sequence $R_{ji}$ through inserting $m$ decoy photon states selected randomly from $\{|+\rangle,|-\rangle,|+y\rangle,|-y\rangle\}$ randomly into $R^{''}_{ji}$. Finally, the new sequence $R_{ji}$ is sent to bidder $i$. Bidders $j$ and $i$ adopt the decoy photon checking technique to guarantee the security of transmission. After that, the leftover qubits are the sequence $R^{''}_{ji}$.

\textbf{Step 6}. Each bidder $j$ sends the sequence $P^{''}_j$ back to Alice. Since Alice knows the initial state of $P_j$, she will measure every qubit in $P^{''}_j$ by utilizing corresponding MB. If the state of qubit has changed, the information is 1, otherwise the information is 0. Obviously the winner can be determined according to the highest bid. Alice announces the highest bid and candidate winner.

\textbf{Step 7}. For resisting Alice's maliciousness, the winner (for example the $j$-th bidder) announces the permutation operator $\Pi^{j}_{(m)}$. Then the other bidders recover the correct orders of $R^{'}_{ji}$ using $\Pi^{j}_{(m)}$ and perform the Bell measurement on the corresponding EPR pairs in the sequence. The fairness of the auction will be guaranteed when all the measurement results are the same as the bid which Alice announced, otherwise the auction is considered to be unfair.

As described above, the whole procedures can be divided into two parts: Steps 1-4 and 6 introduce the auction process among the bidders and the auctioneer; and the post-confirmation process is given in Steps 5 and 7. In next section, we will give an example of the three-party QSA protocol.

\section{An example: three-party quantum sealed-bid auction}
\label{sec:3}
For the sake of clearness, an example is given to explicitly describe a three-party quantum sealed-bid protocol based on single photons. Suppose the auctioneer is Alice and there are two seller agents, Bob and Charlie. And Bob's bid is 1011, Charlie's bid is 0111. For simplicity, we skip the decoy photon checking process on quantum channel; the major process consists of the following five steps (also shown in Fig. 1):

\begin{figure}
  \centering
  \includegraphics[width=11cm]{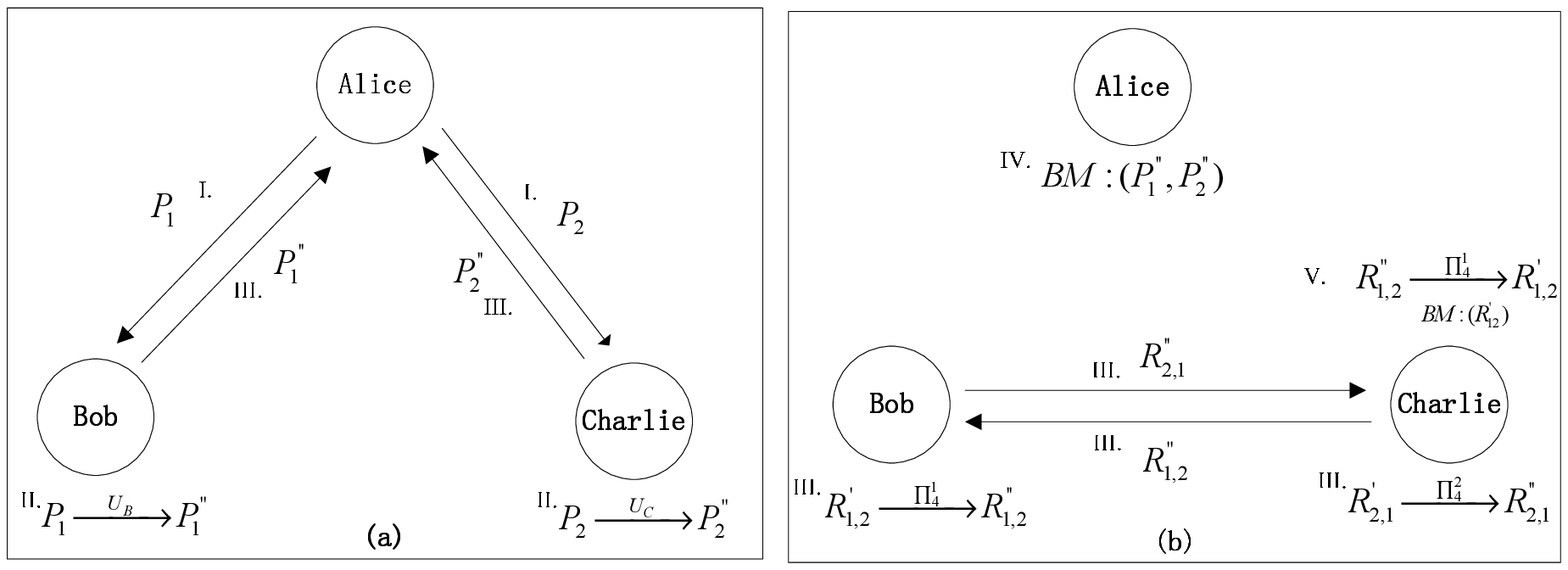}
  \caption{ The process of three-party quantum sealed-bid auction protocol. (a) Bob (Charlie) encrypts his or her bid into the single photons $P_1$ ($P_2$) prepared by Alice, and sends them back to Alice, here, $U_B,U_C\in\{I,i\sigma_y\}$, $P_1\xlongrightarrow{U_B}P^{'}_1$ represents Bob applies the unitary operation $U_B$ on the particle sequence $P_1$ to form $P^{''}_1$, so does $P_2\xlongrightarrow{U_C}P^{'}_2$. (b) The post-confirmation mechanism based on EPR pairs by adding a permutation operator is executed, here, $\Pi^{1}_4$ denotes a permutation operator, $R^{'}_{1,2}\xlongrightarrow{\Pi^{1}_4}R^{''}_{1,2}$ represents Bob performs $\Pi^{1}_4$ on EPR sequence $R^{'}_{1,2}$ to form $R^{''}_{1,2}$, so do $R^{''}_{1,2}\xlongrightarrow{\Pi^{1}_4}R^{'}_{1,2}$ and $R^{'}_{2,1}\xlongrightarrow{\Pi^{2}_4}R^{''}_{2,1}$, and BM represents measurement basis.} \label{Fig. 1}
\end{figure}

\textbf{Step I}. Alice generates 2 groups 4-length qubits randomly in one of $\{|0\rangle,|1\rangle,|+\rangle,|-\rangle\}$, i.e., $P_1=|0\rangle|1\rangle|+\rangle|-\rangle, P_2=|+\rangle|0\rangle|-\rangle|1\rangle$. Then Alice sends $P_1$ to Bob and $P_2$ to Charlie.

\textbf{Step II}: Bob and Charlie encrypt their bids on the respective sequence of quantum particles according to the encoding rule. After that, $P_1$ is transformed into the sequence $P^{''}_1=|1\rangle|1\rangle|-\rangle|+\rangle$ and $P_2$ to $P^{''}_2=|+\rangle|1\rangle|+\rangle|0\rangle$.

\textbf{Step III}. Bob (Charlie) prepares an ordered sequence of EPR pairs $R^{'}_{1,2}=|\varphi^{+}\rangle_{12}|\varphi^{-}\rangle_{34}$ ($R^{'}_{2,1}=|\psi^{-}\rangle_{12}|\varphi^{-}\rangle_{34}$) according to his (her) secret bid, and applies a permutation operator $\Pi^{1}_{(4)}$ ($\Pi^{2}_{(4)}$) on the sequence $R^{'}_{1,2}$ ($R^{'}_{2,1}$) to form the sequence $R^{''}_{1,2}$ ($R^{''}_{2,1}$). The order of sequence $R^{''}_{1,2}$ ($R^{''}_{2,1}$) is 1324 (4123). Bob (Charlie) sends $R^{''}_{1,2}$ ($R^{''}_{2,1}$) to Charlie (Bob) and $P^{''}_1$ ($P^{''}_2$) to Alice.

\textbf{Step IV}. After Alice received $P^{''}_1$ ($P^{''}_2$), she chooses the corresponding MB ($\{|0\rangle,|1\rangle\}$ or $\{|+\rangle,|-\rangle\}$) to measure every qubit in $P^{''}_1$ ($P^{''}_2$). Then she compares the measurement results with the initial states, which follows the rule: if they are different, the corresponding binary bit is 1; otherwise, the binary bit is 0. So she gets two binary sequences, i.e., 1110 (Bob's bid) and 0100 (Charlie's bid). Obviously, Bob's bid is higher than Charlie's. Then Alice will announce that the winner is Bob, and the highest bid is 1110.

\textbf{Step V}. Since the auctioneer Alice announces the result, other bidders will perform a post-confirmation mechanism to verify the reality of the winner. At first, Bob tells Charlie $\Pi^{1}_{(4)}$, then Charlie applies it on the disordered sequence $R^{''}_{1,2}$ and gets the original ordered sequence: $R^{'}_{1,2}=\Pi^{1}_{(4)}R^{''}_{1,2}$, i.e., $|\varphi^{+}\rangle_{12}|\varphi^{-}\rangle_{34}$. After that, she performs the Bell measurement on the EPR pairs sequence $R^{'}_{1,2}$, then she gets Bob's bid, i.e., 1011. Since the bid obtained by Charlie is the same as that one announced by Alice, so we can say the auction is fair and effective.

Up to now, the three-party example which describes the present MQSA protocol is introduced completely. The security analysis for the new protocol will be given in the next section.

\section{Security and efficiency analysis}
\label{sec:4}
\subsection{Security analysis}
\label{sec:4.1}
The crucial issue of any quantum sealed-bid auction is its security. It must guarantee the privacy of every bid's message and fairness of the auction among the bidders. To be specific, it must follows the rules [30], (i) the outer eavesdroppers cannot get any information about bids; (ii) any bidder or a group of bidders cannot obtain bid message from other bidders; (iii) a malicious bidder may collude with the dishonest auctioneer but cannot win over all the other bidders.
\begin{flushleft}
\textbf{(1) Outsider attack}
\end{flushleft}

Similar to Refs. [27,28,30], the decoy photon checking technique is adopted to resist the outer eavesdroppers' attacks, such as the intercept-and-resend attack, disturbance attack and the attack with fake entangled particles, which is proven to be effective to prevent outer eavesdroppers from stealing the bids. Here, we take the intercept-and-resend attack as example to check the security against outsider eavesdroppers' attacks. In the present protocol, the eavesdropper Eve may launch this attack during the process Alice sends $P^{'}_j$ to every bidder $j$ in Step 2 and/or the converse process every bidder $j$ sends the sequence $P^{''}_j$ back to Alice in Step 6. As described in Step 2, Alice inserts the decoy photons, namely detection photons,into $P_j$ and gets new sequence $P^{'}_j$, then sends $P^{'}_j$ to every bidder. At this time, Eve may intercept the transmitted sequence $P^{'}_j$ and make some measurements on them, and tries to steal the original states of $P_j$ Alice prepared. For avoiding being detected by the participants, she will continually send the measured $P^{'}_j$ or new single photon sequence to each bidder. Here, since Eve does not know the positions of detection photons in $P^{'}_j$, she cannot differentiate the detection photons and $P_j$. If she make some measurements on $P^{'}_j$, it will inevitably increase the error rate in Step 3, that means the eavesdropping will be found. Secondly, $P_j$ is prepared randomly from $\{|0\rangle,|1\rangle,|+\rangle,|-\rangle\}$ and the measurement bases are unknown for Eve, she cannot get any information through measurement operations. According the description in Step 5 and 6, the converse process is similar to the above process. so, we can say
the present protocol is secure against the intercept-and-resend attack.

\begin{flushleft}
\textbf{(2) Insider attack}
\end{flushleft}

Suppose a malicious bidder may launch CNOT attack [21,22] to obtain the bid of the $j$-th bidder. Before auction begins, she (he) prepares a set of auxiliary particles $S_e$. In Step 2, She (He) will intercept the quantum sequence $P_j$ sent by Alice, perform a CNOT operation on every particle in $P_j$ and the corresponding auxiliary particle in $S_e$ and resend $P_j$ to bidder $j$. When in Step 3 Alice and bidder $j$ perform the eavesdropper checking of quantum channel,it is easy to found that the states of checking qubits in $P_j$ are disturbed. Then the protocol will be abandoned. For example, auxiliary particle is $|0\rangle_e$, and checking qubits are generated randomly in one of $\{+\rangle,|-\rangle,|+y\rangle,|-y\rangle\}$. The relationship between the initial composite system, which consists of the checking qubit and the ancilla, and the one after CNOT operation is listed in Table 1.
\begin{table}[h]\normalsize
\caption{The relationship between the initial composite system and the one after CNOT operation}
\newcommand{\tabincell}[2]{\begin{tabular}{@{}#1@{}}#2\end{tabular}}
\begin{tabular}{|l|l|l|l|}
  \hline
  \tabincell{c}{Auxiliary\\ qubit}&\tabincell{c}{Checking\\ qubit}&Composite sytem(initial)&Composite system(after  CNOT)\\
   \hline
  $|0\rangle_e$ & $|+\rangle$ &
  \begin{minipage}{1.8in}
  \begin{equation*}
  \begin{aligned}
    |{\bigwedge}^{1}\rangle &=|+\rangle_c|0\rangle_e\\
                        &=\frac{1}{\sqrt{2}}(|00\rangle_{ce}+|10\rangle_{ce})
  \end{aligned}
  \end{equation*}
  \end{minipage}
   &
  \begin{minipage}{1.8in}
  \begin{equation*}
  \begin{aligned}
     |{\bigwedge}^{2}\rangle &=\frac{1}{\sqrt{2}}(|00\rangle_{ce}+|11\rangle_{ce})\\
                  &=\frac{1}{\sqrt{2}}(|++\rangle_{ce}+|--\rangle_{ce})
  \end{aligned}
  \end{equation*}
  \end{minipage}
  \\
   \hline
  $|0\rangle_e$ & $|-\rangle$ &
  \begin{minipage}{1.8in}
  \begin{equation*}
  \begin{aligned}
    |{\bigwedge}^{1}\rangle &=|-\rangle_c|0\rangle_e\\
                        &=\frac{1}{\sqrt{2}}(|00\rangle_{ce}-|10\rangle_{ce})
 \end{aligned}
  \end{equation*}
  \end{minipage}
  &
  \begin{minipage}{1.8in}
  \begin{equation*}
  \begin{aligned}
     |{\bigwedge}^{2}\rangle &=\frac{1}{\sqrt{2}}(|00\rangle_{ce}-|11\rangle_{ce})\\
                  &=\frac{1}{\sqrt{2}}(|+-\rangle_{ce}+|-+\rangle_{ce})
  \end{aligned}
  \end{equation*}
  \end{minipage}
  \\
   \hline
  $|0\rangle_e$& $|+y\rangle$ &
  \begin{minipage}{1.8in}
  \begin{equation*}
  \begin{aligned}
    |{\bigwedge}^{1}\rangle &=|+y\rangle_c|0\rangle_e\\
                        &=\frac{1}{\sqrt{2}}(|00\rangle_{ce}+i|10\rangle_{ce})
  \end{aligned}
  \end{equation*}
  \end{minipage}
  &
  \begin{minipage}{2.3in}
  \begin{equation*}
  \begin{aligned}
    |{\bigwedge}^{2}\rangle &=\frac{1}{\sqrt{2}}(|00\rangle_{ce}+i|11\rangle_{ce})\\
                        &=\frac{1}{\sqrt{2}}(|+y\rangle_{c}|+\rangle_e+|-y\rangle_{c}|-\rangle_e)
  \end{aligned}
  \end{equation*}
  \end{minipage}
  \\
   \hline
  $|0\rangle_e$ & $|-y\rangle$ &
  \begin{minipage}{1.8in}
  \begin{equation*}
  \begin{aligned}
    |{\bigwedge}^{1}\rangle &=|-y\rangle_c|0\rangle_e\\
                        &=\frac{1}{\sqrt{2}}(|00\rangle_{ce}-i|10\rangle_{ce})
  \end{aligned}
  \end{equation*}
  \end{minipage}
  &
   \begin{minipage}{2.3in}
  \begin{equation*}
  \begin{aligned}
    |{\bigwedge}^{2}\rangle &=\frac{1}{\sqrt{2}}(|00\rangle_{ce}-i|11\rangle_{ce})\\
                        &=\frac{1}{\sqrt{2}}(|+y\rangle_{c}|-\rangle_e+|-y\rangle_{c}|+\rangle_e)
  \end{aligned}
  \end{equation*}
  \end{minipage}
   \\
  \hline
\end{tabular}
\end{table}
From Table 1, we can see the states of the checking qubits are disturbed, so this CNOT attack can be detected. By now, we have proved that the malicious bidder cannot get other's bid without introducing any error.

\begin{flushleft}
\textbf{(3) collusion attack}
\end{flushleft}

There are two types of collusion attack: (1) auctioneer Alice may collude with a malicious bidder and help him/her win this auction; (2) many bidders collude with each other to get the bid of the honest bidder. If the auctioneer wants to help a malicious bidder to win the auction successfully, he may falsely claim the winner is that malicious bidder under the case that the winner is one of others in Step 6. But it must be noted that Step 5 and Step 7 introduce a post-confirmation mechanism to check the reality of the winner. In Step 5, all bidders prepare the sequences of EPR pairs according to their bids and send them to others. In Step 7, after auctioneer announces the winner is the $j$-th bidder, other bidders can measure EPR pairs in sequence $R^{'}_{ji}$ sent by the winner. If the measurement result is different from that announced by Alice, obviously, the collusion of the dishonest bidder and the malicious auctioneer can be found, then, this auction is abandoned. Now we consider another collusion attack among the dishonest bidders. For launching collusion attack successfully, the dishonest bidders may try to measure EPR pairs in $R^{'}_{ji}$ to get the $j$-th bidder's bid (suppose he or she is honest) after checking the security of the channel in Step 5. However, since they are disordered by the permutation operator before the sequences of EPR pairs $R^{'}_{ji}$ are sent to other bidders in Step 5, the correct positions of EPR corresponding pairs are unknown to anyone except the sender. Under this circumstance, any bidder cannot get sender's bid through measuring EPR pairs. So our protocol also can resist against the collusion among several dishonest bidders to get other bids.

\begin{flushleft}
\textbf{(4) Side channel attack}
\end{flushleft}

Our protocol has been considered and discussed in the idea settings. In practical scenario, the quantum protocol inevitably suffers from some side channel attacks. Since our protocol transmits the same photons more than once, it may suffer from the Trojan horse attacks(also called invisible photon attacks), and such kind of quantum transmission has been discussed [32,33]. To prevent this type of attacks, the participants can install a special quantum optical device such as the wavelength quantum filter and the photon number splitters (PNS) to detect an attack. According to Refs. [32, 33], Eve's invisible photons can be filtered out by using the wavelength quantum filter, and the PNS can split each legitimate photon to discover the delay photons. If there is an irrational high rate of multi-photon signal, then the attack can be detected. As pointed out in Ref. [32], this kind of Trojan horse attack is not an exploit of a weakness of the protocol in itself, but rather an exploit of a weakness in certain imperfect implementations. Without the imperfection of the single-photon detectors, this kind of Trojan horse attack will not exist any longer.

\subsection{Efficiency analysis}
\label{sec:4.2}
As analyzed above, the present MQSA protocol suffices for three security requirements [30]. That is to say, it guarantees the privacy of every bid's message and fairness of the auction among the bidders.

In addition, the efficiency is another outstanding advantage in our MQSA protocol. In Refs. [20-28], the GHZ states are utilized as the bid message carrier, where every 3-qubit GHZ state can transmit two classic bits (cbits) bid message. Taking the quantum consumption rate $\xi$ (the amount of qubits every transmitted cbit bid message) into account, the quantum consumption rate in Refs. [20-23,25-28] is $\xi_{GHZ}=3/2=1.5$. While in Refs. [29, 30], one EPR pairs (2-qubit) is employed to transmit two cbits bid message, so the rate $\xi_{EPR}=2/2=1.0$. And in our MQSA protocol, one single photon (1-qubit) is used to transmit one cbit bid message, so $\xi_{Our}=1/1=1.0$. The comparison among these three kinds of QSA/MQSA protocols is listed in Table 2.
\begin{table}[h]\normalsize
\caption{The comparison among different kinds of QSA or MQSA protcols}
\newcommand{\tabincell}[2]{\begin{tabular}{@{}#1@{}}#2\end{tabular}}
\begin{tabular}{|l|l|l|l|l|}
  \hline
   \tabincell{c}{protocol} & \tabincell{c}{quantum resource \\(qubit)} & \tabincell{c}{bid message \\(cbit)} & \tabincell{c}{quantum consumption \\rate $\xi$ (qubit/cbit)} & \tabincell{c}{detection quantum \\state } \\
  \hline
  Refs.[20-23,25-28] & \tabincell{c}{ GHZ \\(3 qubits)} & 2 cbits & 3/2=1.5 &  \tabincell{c} { GHZ$^{[20-22]}$, \\EPR pairs$^{[23]}$, \\single photon$^{[25-28]}$}\\
  \hline
  Refs.[29,30] & \tabincell{c}{EPR pairs \\(2 qubits)} & 2 cbits & 2/2=1.0 & \tabincell{c} { EPR pairs$^{[29]}$ , \\single photon$^{[30]}$}\\
  \hline
  our protocol & \tabincell{c}{single photon \\(1 qubit)} & 1 cbit & 1/1=1.0 & single photon\\
  \hline
\end{tabular}
\end{table}
As shown in the table, the quantum consumption of our protocol is the same as the protocols [29, 30] using EPR pairs, but obviously lower than the protocols [20-23,25-28] using GHZ states. On the other way, taking the detection quantum state into account, we can find our protocol takes the single photon as the detection quantum state as the same as Refs.[25-28,30], which saves more qubits than Refs.[23, 29] using EPR pairs and Refs.[20-22] using GHZ state (shown in the last column of Table 2).

As we all know, the single photon is viewed to be easier to be prepared in the current technology than the entangled states, such as EPR pairs, the GHZ state. Since we employ the single photons as the message carrier, it shows our protocol is more feasible and economic than those protocols[20-23,25-30] based on entangled states.

\section{Discussion and conclusion}
\label{sec:5}
In conclusion, different from those protocols using entangled states as quantum resource, we proposed a novel MQSA protocol using single photons as message carrier. In the protocol, the decoy photon technique is used to defeat against outsider attack, insider attack and collusion attack, while an improved post-confirmation mechanism employing EPR pairs and a permutation operator is introduced to guarantee the fairness of auction. Compared with other QSA or MQSA protocols, our protocol can save the quantum consumption, and shows more efficient as well as feasible in the current technology.

It should be noted that the EPR pairs are still used in the post-confirmation process while the single photons are utilized as the bid carrier to implement the process of the auction. There are some interesting open problem to be solved. Is the entangled state necessary in the post-confirmation mechanism? And is there any other kind of feasible method to guarantee the fairness instead of the post-confirmation mechanism? On the other hand, how to generalize our multiparty quantum sealed-bid auction into the random \emph{d}-dimension scenario is our another work to be further studied in the future.

\begin{acknowledgements}
This work is supported by the National Nature Science Foundation of China (Grant Nos. 61502101, 61501247, 61373016 and 61373131), the Priority Academic Program Development of Jiangsu Higher Education Institutions (PAPD), the Natural Science Foundation of Jiangsu Province under Grant No. BK20140651, and the Guangxi Science Fund for Distinguished Young Scholars (Grant No. 2015GXNSFGA139001).
\end{acknowledgements}



\end{document}